\documentclass[fleqn,10pt]{wlscirep}
\usepackage[utf8]{inputenc}
\usepackage[T1]{fontenc}
\usepackage{bm}
\usepackage{algorithm2e}
\usepackage{newtxmath,newtxtext}
\DeclareMathAlphabet{\mathcal}{OMS}{cmsy}{m}{n}

\title{Stochastic approach for quantum metrology with generic Hamiltonians}

\author[1,2,*]{Le Bin Ho}
\affil[1]{Frontier Research Institute for Interdisciplinary Sciences, Tohoku University, Sendai 980-8578, Japan}
\affil[2]{Department of Applied Physics, Graduate School of Engineering, Tohoku University, Sendai 980-8579, Japan}

\affil[*]{binho@fris.tohoku.ac.jp}

\begin{abstract}
Recently, variational quantum metrology 
was proposed for Hamiltonians with multiplicative parameters, 
wherein
the estimation precision 
can be
optimized
via variational circuits. 
However, systems with generic Hamiltonians
still lack these variational schemes. 
This work introduces a quantum-circuit-based 
approach for studying quantum metrology 
with generic Hamiltonians. We present a time-dependent 
stochastic parameter-shift rule for the derivatives of 
evolved quantum states,
whereby the quantum Fisher information 
can be obtained. 
The scheme can be executed in universal 
quantum computers under 
the family of parameterized gates. 
In magnetic field estimations, 
we demonstrate the consistency between the results obtained 
from the stochastic parameter-shift rule and the exact results, 
while the results obtained from a standard parameter-shift rule 
slightly deviate from the exact ones. 
Our work sheds light on 
studying quantum metrology  
with generic Hamiltonians 
using quantum circuit algorithms. 

\end{abstract}
\begin{document}

\flushbottom
\twocolumn
\sloppy
\maketitle
%
%


\section*{Introduction}
The objective of quantum metrology is 
that using nonclassical quantum resources
to enhance the precision in 
the estimation of unknown parameters
\cite{Chalopin2018,RevModPhys.90.035005},
including entanglement \cite{PhysRevLett.102.100401,
PhysRevLett.79.3865,doi:10.1126/science.1104149,
PhysRevLett.96.010401} and 
squeezing states \cite{PhysRevA.46.R6797,PhysRevA.50.67}.
It has a wide range of applications,
from quantum magnetometry
\cite{PhysRevA.102.022602,PhysRevLett.125.020501},
to quantum clocks \cite{RevModPhys.83.331,RevModPhys.87.637},
quantum imaging \cite{Lugiato_2002,Moreau2019},
gravitational-wave detection
\cite{Schnabel2010}, 
dark matter detection 
\cite{10.1007/978-3-030-31593-1_5},
and so on.
So far, it was demonstrated 
quantum-enhanced beyond the Heisenberg limit
with nonlinear interaction 
\cite{PhysRevLett.102.100401,PhysRevX.2.041006,PhysRevA.89.022107},
interaction-based \cite{Napolitano2011,PhysRevLett.119.193601},
and even without entanglement \cite{RevModPhys.90.035006}.
Quantum metrology 
under noisy environments
\cite{PhysRevLett.112.120405,Tsang_2013,Haa},
post-selection measurements
\cite{PhysRevLett.114.210801,
Arvidsson-Shukur2020,doi:10.1063/5.0024555}, 
and quantum error correction
\cite{PhysRevLett.112.150802,
Gorecki2020optimalprobeserror,
PhysRevResearch.2.013235}
are also extensively reported.

The cornerstone of quantum metrology 
is the quantum estimation theory, 
which imposes the 
lower bound of
precision 
by 
the quantum Cram{\'e}r-Rao inequality 
\cite{doi:10.1142/S0219749909004839}. 
The quantum Cram{\'e}r-Rao bound
is associated with 
quantum Fisher information 
(QFI) for single-parameter estimation
and quantum Fisher information matrix 
(QFIM) for multiparameter estimation.

Numerous 
studies on quantum metrology 
mainly focus on multiplicative 
parameters of Hamiltonians, 
e.g., a parameter $\theta$
in a Hamiltonian $\theta H$
\cite{PhysRevA.90.022117}.
However, recent attention was raised
to generic parameters of Hamiltonians, 
such as quantum magnetometry 
\cite{PhysRevA.90.022117,
PhysRevLett.116.030801,
PhysRevA.102.022602},
unitary parametrization process 
\cite{Liu2015,PhysRevA.92.012312},
and time-dependent Hamiltonians 
\cite{PhysRevLett.115.110401,Pang2017}.
While the estimation with generic Hamiltonians 
shares some typical 
properties 
with the multiplicative case, 
it likewise indicates other distinct features, such as 
getting high efficiency with time scaling
\cite{PhysRevA.90.022117} 
and quantum control 
\cite{PhysRevLett.115.110401,Pang2017}.
The study of 
quantum bounds 
in these generic 
cases
will open a broad range
of potential applications in quantum metrology.

On the other side, quantum computers can outperform 
classical ones and open significant quantum advantages for 
exponentially speeding up various computational tasks 
\cite{Arute2019,ChinaSupremacy}.
Specifically, using Noisy Intermediate-Scale Quantum
computers \cite{Preskill2018quantumcomputingin}
resulted in the brilliant growth 
of different quantum algorithms
(see Refs. \cite{Cerezo2021,Montanaro2016}.)
Among them, variational 
quantum algorithms \cite{Cerezo2021}
are the most 
promising approach
for improving the efficiency 
in noisy and few-qubits devices.
These algorithms include
variational quantum eigensolvers 
\cite{Peruzzo2014,PhysRevResearch.1.033062,
PRXQuantum.2.020337},
quantum approximate optimization algorithms
\cite{PhysRevX.10.021067},
new frontiers in quantum foundations 
\cite{Arrasmith2019,Koczor_2020,Meyer2021},
and so on. 

Besides, many computational tools 
based on variational quantum circuits 
were developed, including the 
standard parameter-shirt rules 
(Stand.PSR)
\cite{PhysRevA.98.032309,PhysRevA.99.032331} 
and quantum natural gradient
\cite{Stokes2020quantumnatural}.
The Stand.PSR allows us 
to get the exact partial derivatives
of any function by calculating it
with different shifted parameters in the circuits.
%
However, it only applies to cases 
where the gate's generators commute.
Otherwise, to apply the Stand.PSR, 
additional treatments are required, such as
Hamiltonian simulation techniques \cite{Andrew}.
Recently, Banchi and Crooks in their seminal work,
have developed a stochastic parameter-shift rule (Stoc.PSR) 
for general quantum evolutions,
which relies on the stochastic repetitions of
quantum measurement
\cite{Banchi2021measuringanalytic}. 

So far, different variational quantum algorithms
for quantum metrology
were developed,
which open a new way to achieve
quantum-enhanced precision
\cite{PhysRevLett.123.260505,Koczor_2020,Meyer2021}. 
Moreover, the Stand.PSR was 
widely used in various 
aspects, including finding the QFI 
with multiplicative
Hamiltonians 
\cite{Meyer2021fisherinformationin,PhysRevResearch.4.013083}.
However, it is lacking in the study of generic Hamiltonians.
In reality, many systems are governed by generic
Hamiltonians. Therefore, 
studying
these cases
using quantum algorithms is urgent. 

This paper 
introduces 
a general time-dependent Stoc.PSR and 
applies it to
quantum metrology. %
We utilize the proposed Stoc.PSR for 
the derivatives of
evolved quantum states, 
then compute the QFIM (or QFI)
and examine the estimation precision. 
Our scheme can execute 
in universal quantum computers 
under the family of 
parameterized gates. 
In magnetic field estimations, 
we show an excellent agreement 
between the results obtained from the
Stoc.PSR and the exact results
while the Stand.PSR's results deviate 
from the exact values.
This observation suggests the 
significance of the Stoc.PSR 
for studying quantum metrology 
with generic Hamiltonians
and its applicability to variational 
quantum metrology.
Furthermore, we extend our approach 
to examine the precision in many-body 
Hamiltonian tomography,
such as estimating unknown 
coupling constants in the Hamiltonian.

\section*{Results}
\subsection*{Quantum parameters 
estimation for generic Hamiltonians}
Estimation is a measurement process 
that uses 
a probe to extract information 
from an interesting system with 
$d$ unknown parameters in a field
$\bm B = \phi_1\bm e_1 
+ \cdots + \phi_d\bm e_d$,
where $\{\bm e_j\}$ are unit vectors in $\{j\}$ directions.
The probe interacts with the system 
through a generic Hamiltonian
$H(\bm\phi) = \bm B \cdot \bm H 
= \sum_{j=1}^d \phi_jH_j$,
where the $\{H_j\}$ do not necessarily commute.
The task of quantum parameters estimation 
is to evaluate these unknown 
coefficients by measuring the probe.


Let $\rho_0$ be the initial probe state,
it evolves to  
$\rho(\bm\phi) = U(\bm\phi)\rho_0U^\dagger(\bm\phi)$
after the interaction,
where $U(\bm\phi) = e^{-itH(\bm\phi)}$ is the unitary 
evolution during the interaction time $t$.
Note that $H(\bm\phi)$ is a general Hamiltonian, 
therefore $U(\bm\phi)$ cannot be expanded 
in terms of multiplicative.
By measuring the probe state in a general basis set, 
such as the positive operator-valued measure (POVM) 
${E_x}$ for the outcome $x$, 
one can obtain the corresponding probability 
distribution $p(x|\bm\phi) = 
{\rm tr}\big[\rho(\bm\phi)E_x\big]$, 
which can be used to estimate 
the unknown parameters $\bm\phi$.

In the estimation theory, different estimators can be used to obtain the estimated value $\bm{\check{\phi}}(x)$ of the unknown parameters $\bm\phi$, each yielding different precisions. The precision is characterized by the covariance matrix $C(\bm\phi) = {\rm E}\big[(\bm\phi - {\rm E}[\bm{\check{\phi}}(x)]) (\bm\phi - {\rm E}[\bm{\check{\phi}}(x)])^\intercal\big]$ \cite{doi:10.1063/5.0024555}, where E$[\bm{\check{X}}]=\int p(x|\bm X)
\bm{\check{X}}(x) {\rm d}x$ is the expectation value
of the estimator $\bm{\check{X}}(x)$.
The diagonal term $C_{k,k} \equiv 
\Delta^2\phi_k = {\rm E}[\phi_k^2]-{\rm E}^2[\phi_k]$
is the variance for estimating $\phi_k$, and 
the off-diagonal term $C_{k,l}$ is the covariance 
between $\phi_k$ and $\phi_l$.
An estimator is unbiased when 
E$[\check{\phi}_k(x)] = \phi_k\;,\ 
\forall k \in \{ 1,\cdots, d\}$. 
The precision 
obeys
classical and quantum Cram{\'e}r-Rao bounds (CRBs)
 \cite{doi:10.1142/S0219749909004839}
\begin{align}\label{eq:CRB}
M\cdot C(\bm\phi) \ge F^{-1}(\bm\phi) \ge Q^{-1}(\bm\phi),
\end{align}
where $M$ is the number of repeated measurements, 
$F(\bm\phi)$ is the classical Fisher 
information matrix (CFIM) defined by
\begin{align}\label{eq:F}
F_{k,l} = \int \dfrac{1}{p(x|\bm\phi)}
\big[\partial_{\phi_k} p(x|\bm\phi)\big]
\big[\partial_{\phi_l} p(x|\bm\phi)\big]
{\rm d}x,
\end{align}
and the maximum over all possible 
measurements $\{E_x\}$ yields the quantum
Fisher information matrix (QFIM) $Q(\bm\phi)$ 
with elements
\begin{align}\label{eq:Q}
    Q_{k,l} = \dfrac{1}{2}{\rm tr}\big[
    \rho(\bm\phi)\{L_k,L_l\}
    \big],
\end{align}
where $L_k$ is
the symmetric logarithmic derivative (SLD)
that obeys $2\partial_{\phi_k}\rho(\bm\phi) = 
L_k\rho(\bm\phi) + \rho(\bm\phi)L_k$
\cite{doi:10.1142/S0219749909004839}. 
For a single parameter estimation (such as $\phi$), 
the CRBs simplify to $\Delta^2\phi 
\ge 1/F(\phi) \ge 1/Q(\phi)$, 
where $F(\phi) = \int p(x|\phi)
\big[\partial_\phi\ln p(x|\phi)\big]^2{\rm d}x$
and $Q(\phi) = {\rm tr}\big[L^2\ \rho(\phi)\big]$
are the classical 
and quantum Fisher information, 
respectively. 
Note that both CFIM and QFIM may 
depend on the parameters $\bm\phi$ 
regardless of the unitary process.

The QFI and QFIM set ultimate bounds for 
the estimation precision of any estimator. 
Therefore, it is crucial to derive
these QFI and QFIM for the 
estimation theory with generic 
Hamiltonians. 
Let us start with the derivative of the unitary evolution
\cite{PhysRevLett.116.030801,doi:10.1063/1.1705306}
\begin{align}\label{eq:dU}
    \notag \dfrac{\partial e^{-itH(\bm\phi)}}
    {\partial\phi_j} &= -i\int_0^t
    e^{-i(t-s)H(\bm\phi)}
    \big[\partial_{\phi_j} H(\bm\phi)\big]
    e^{-isH(\bm\phi)}
    {\rm d}s, \\ 
    &=-iU(\bm\phi)Y_j,
\end{align}
where 
$
    Y_j = 
     \int_0^t
    e^{isH(\bm\phi)}\big[\partial_{\phi_j} 
    H(\bm\phi)\big]e^{-isH(\bm\phi)}
    {\rm d}s
$ is a Hermitian operator
\cite{PhysRevLett.116.030801}.
Then, we obtain 
\begin{align}\label{eq:drho}
    \dfrac{\partial \rho(\bm\phi)}
    {\partial\phi_j} = -iU(\bm\phi)
    \big[Y_j,\rho_0\big]U^\dagger(\bm\phi).
\end{align}
The QFIM \eqref{eq:Q} 
yields
\begin{align}\label{eq:iQ}
   Q_{k,l} = 2\sum_{p_\lambda+p_{\lambda'}>0}
   \dfrac{\langle \lambda|\partial_{\phi_k}\rho(\bm\phi)|\lambda'\rangle
   \langle \lambda'|\partial_{\phi_l}\rho(\bm\phi)|\lambda\rangle}
   {p_\lambda+p_{\lambda'}},
\end{align}
for $\rho(\bm\phi) = \sum_\lambda 
p_\lambda|\lambda\rangle\langle \lambda|$, 
and $\partial_{\phi_{k}}\rho(\bm\phi)$ is given 
from Eq.~\eqref{eq:drho}.
%
For pure quantum states, i.e.,
$\rho_0 = |\psi_0
\rangle\langle\psi_0|$, 
the QFIM is defined by \cite{doi:10.1142/S0219749909004839}
\begin{align}\label{eq:Qpure}
\notag    
    Q_{k,l} &= 4{\rm Re}
    \big[\langle\partial_{\phi_k}\psi(\bm\phi)
    |\partial_{\phi_l}\psi(\bm\phi)\rangle \\
    &\hspace{1.2cm} 
    -\langle\partial_{\phi_k}
            \psi(\bm\phi)|\psi(\bm\phi)\rangle
    \langle\psi(\bm\phi)|\partial_{\phi_l}\psi(\bm\phi)\rangle
    \big],
\end{align}
where $|\psi(\bm\phi)\rangle = U(\bm\phi)|\psi_0\rangle$
is the evolved probe state.
Substituting Eq.~\eqref{eq:dU} 
into Eq.~\eqref{eq:Qpure}, 
it yields
\cite{PhysRevLett.116.030801,PhysRevA.102.022602}
\begin{align}\label{eq:Qpure1}
    Q_{k,l} = 4{\rm Re}
    \big[
    \langle\psi_0|Y_kY_l|\psi_0\rangle
    -\langle\psi_0|Y_k|\psi_0\rangle
    \langle\psi_0|Y_l|\psi_0\rangle
    \big].
\end{align}

Computing QFIM and QFI requires 
the derivatives of the probe state, 
i.e., $\partial_{\phi_j}\rho(\bm\phi),\
\forall j\in\{1,\cdots,d\}$. 
Hereafter, we introduce a stochastic parameter-shift rule
(Stoc.PSR) to compute these derivatives on quantum circuits,
allowing for precision evaluation in different quantum computing platforms.

\subsection*{Stochastic parameter-shift rule}
In this section, we present 
a time-dependent stochastic parameter-shift rule (Stoc.PSR)
for quantum metrology with generic Hamiltonians, where we 
particularly calculate $\partial_{\phi_j}\rho(\bm\phi)$
using quantum circuits. 
This method is thus helpful for 
studying different variational quantum algorithms 
\cite{Cerezo2021}, 
including variational quantum metrology 
\cite{Koczor_2020,Meyer2021} and 
evaluating Fubini-Study metric tensor 
in quantum natural gradient 
\cite{Stokes2020quantumnatural}.

\begin{figure}[t]
\centering
\includegraphics[width=8.6cm]{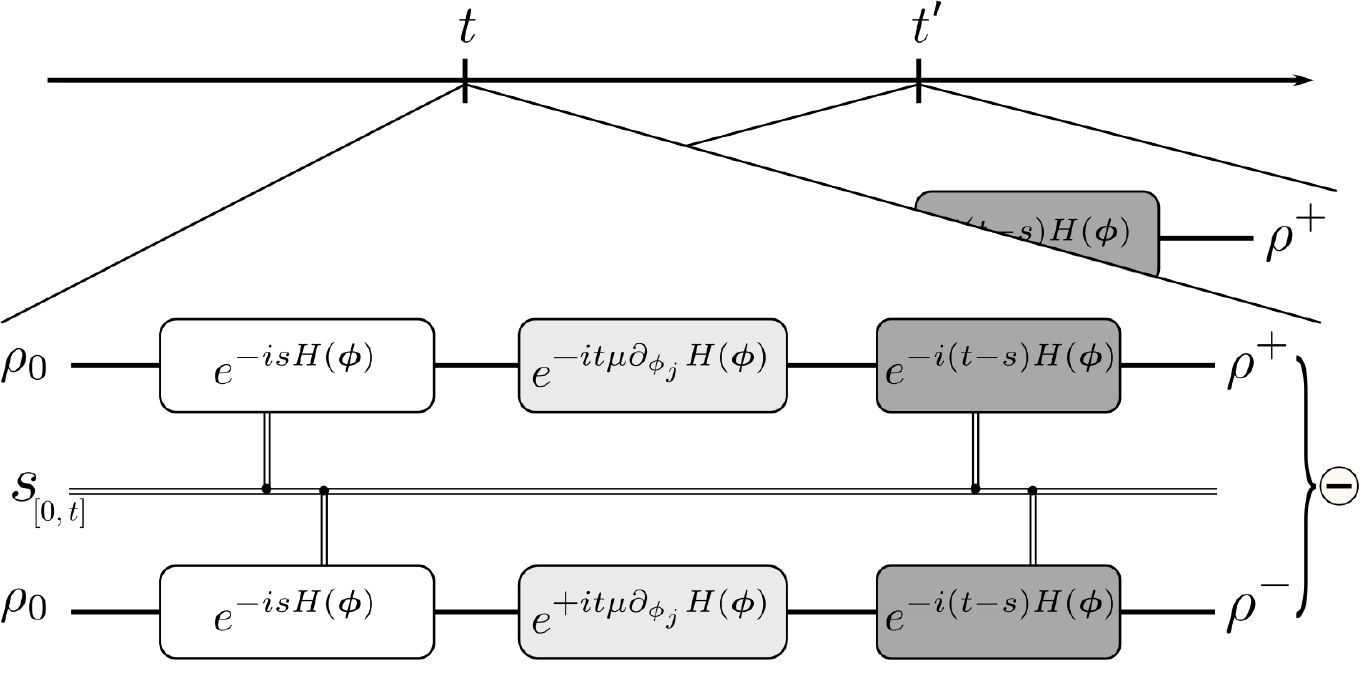}
\caption{\textbf{Quantum circuit for
time-dependent stochastic 
parameter-shift rule (Stoc.PSR).}
For every time $t$, we first prepare
a quantum state $\rho_0$ and 
generate a random number $s \in [0,t]$. 
A sequence of gates $e^{-isH(\bm\phi)}$,
$e^{-it\mu[\partial_{\phi_j}H(\bm\phi)]}$, and 
 $e^{-i(t-s)H(\bm\phi)}$
applies to the circuit and produces $\rho^+$.
We then repeat the scheme while replacing 
$e^{-it\mu[\partial_{\phi_j}H(\bm\phi)]}$
by $e^{it\mu[\partial_{\phi_j}H(\bm\phi)]}$
and compute $\rho^-$.
The derivative $\partial_{\phi_j}\rho(\bm\phi)$ 
is obtained via $\rho^+-\rho^-$.
The process is then repeated for all $\phi_j$
to get the QFIM $Q(\bm\phi)$. Then, we move
to the next time $t'$ and do the same procedure. 
}
\label{fig:1}
\end{figure}

We first recast Eq.~\eqref{eq:drho} in the following form
\begin{align}\label{eq:drho_re}
    \dfrac{\partial \rho(\bm\phi)}
    {\partial\phi_j} = -i\int_0^t U(\bm\phi)
    \big[O_j,\rho_0\big]U^\dagger(\bm\phi) ds,
\end{align}
where $ O_j = e^{isH(\bm\phi)}\big[\partial_{\phi_j} 
    H(\bm\phi)\big]e^{-isH(\bm\phi)}
$.
Referring to \cite{Banchi2021measuringanalytic}
and using the Baker-Campbell-Hausdorff formula 
\cite{Achilles2012} for $H_j^2 = I$, 
we derive
\begin{align}\label{eq:dZ}
    \big[O_j,\rho_0\big]
    = \dfrac{i}{\sin(2t\mu)}
    \Big[e^{-it\mu O_j}\rho_0e^{it\mu O_j} - 
    e^{it\mu O_j}\rho_0e^{-it\mu O_j}\Big],
\end{align}
for all $t\mu \notin \frac{\pi}{2} \mathbb{Z}$.
See Methods section for the detailed proof.
Recall that Ref.~\cite{Banchi2021measuringanalytic}
fixes $\mu = \pi/4$ and $t = 1$.
Here, we consider any time $t$ and introduce $\mu$ 
as an arbitrary parameter shift, 
which makes our scheme more general,
especially in time-dependent and noisy metrology.
For $t\mu = \pi/4$, it reduces to
Ref.~\cite{Banchi2021measuringanalytic}
and maximizes the accuracy
for parameter-shift approaches
(se also Ref.~\cite{PhysRevA.103.012405}.)
Finally, 
using $e^{-it\mu O_j} = e^{isH(\bm\phi)}
e^{-it\mu[\partial_{\phi_j} 
H(\bm\phi)]}
 e^{-isH(\bm\phi)}
$ \cite{doi:10.1063/1.1705306},
and subtituting Eq.~\eqref{eq:dZ}
into Eq.~\eqref{eq:drho_re},
we obtain (see the Methods section): 
\begin{align}\label{eq:drhoFi}
    \dfrac{\partial\rho(\bm\phi)}
    {\partial\phi_j}
    =\dfrac{1}{\sin(2t\mu)}
    \int_0^t\Big[\rho_j^+(\bm\phi,s) 
    - \rho_j^-(\bm\phi,s)\Big] {\rm d}s,
\end{align}
which is the time-dependent
stochastic parameter-shift rule (Stoc.PSR),
where
\begin{align}\label{eq:rhopmUpm}
    \rho_j^\pm(\bm\phi,s)
    &= U_j^\pm(\bm\phi,s)\rho_0 \big[U_j^\pm(\bm\phi)\big]^\dagger, \\
    U_j^\pm(\bm\phi,s)
    &=e^{-i(t-s)H(\bm\phi)}
e^{\mp it\mu \big[\partial_{\phi_j} H(\bm\phi)\big]}
e^{-isH(\bm\phi)}.
\end{align}

The algorithm for time-dependent Stoc.PSR
is described in Algorithm~\ref{al:1},
which is an extended version of the original 
(without time-dependent) in Ref. 
\cite{Banchi2021measuringanalytic}.
%
\begin{algorithm}[h]
    \caption{Stochastic parameter-shift rule for calculating $\partial_{\phi_j}\rho(\bm\phi)$
in quantum circuits.}\label{al:1}
       \KwData{$\rho_0, \bm\phi = (\phi_1,\cdots,\phi_d), 
				H(\bm\phi) = \sum_j\phi_jH_j$}
        \KwResult{$Q(\bm\phi)$}
        $T \gets $ time (array)
        $N \gets $ sampling number
        $\mu \gets $ parameter-shift(rad)
        \For {$t$ in $T$}{
            \For {$j = 1,\cdots,d$}{
               \For {$n = 1,\cdots,N$}{
                     $s$ = random(0, $t$)
                     get $U^{\pm}_j(\bm\phi,s)$
                     get $\rho^{\pm}_j(\bm\phi,s)$
		    get $\partial_j\  +\!= \rho_j^{+}(\bm\phi,s)-\rho_j^{-}(\bm\phi,s)$    
               }
	       $\partial_j$ = $\partial_j*\dfrac{t}{N}\dfrac{1}{\sin(2t\mu)}$	   
	       /* comes from Eq.~\eqref{eq:drhoFi}, 
		where $t/N$ is the average in Monte-Carlo sampling.    */
            }   
            get $Q(\bm\phi)$ /* from Eq.~\eqref{eq:iQ} 
 			 or \eqref{eq:Qpure}. */              
           }
 \end{algorithm}
Figure~\ref{fig:1} depicts a quantum circuit 
for the Stoc.PSR. 
To obtain $\partial_{\phi_j}\rho(\bm\phi)$ for a given time $t$, we perform the following steps: (s1) generate a random number $s$ from a normal distribution within the interval $[0,t]$; (s2) initialize the circuit with $\rho_0$; (s3) apply the quantum gates $e^{-isH(\bm\phi)}$, $e^{-it\mu [\partial_{\phi_j}H(\bm\phi)]}$, and $e^{-i(t-s)H(\bm\phi)}$; (s4) extract the final state $\rho^+$ from the circuit; (s5) repeat steps s2-s4, replacing $e^{-it\mu [\partial_{\phi_j}H(\bm\phi)]}$ with $e^{it\mu [\partial_{\phi_j}H(\bm\phi)]}$, and assign the quantum state to $\rho^-$; (s6) repeat steps s1-s5 $N$ times and compute the derivative via $\frac{t}{N*\sin(2t\mu)}\sum_{n=1}^N(\rho^+-\rho^-)$. 
The term $t/N$ comes from Monte-Carlo sampling, 
i.e., $\int_a^b f(x) dx \approx \frac{b-a}{N}\sum_{i=1}^Nf(x_i)$. 
Apply the procedure for all $j\in {1,\cdots, d}$ and use Eqs. (\ref{eq:iQ}-\ref{eq:Qpure}) we can compute the QFIM. 
Finally, we repeat the scheme for other time instances.

Note that
the scheme
can be implemented in universal quantum computers.
Assuming a programmable quantum computer 
that can execute a family 
of native quantum gates $U(t,\bm\phi) 
= e^{-itH(\bm\phi)}$, where $H(\bm\phi) 
= \sum_j\phi_jH_j$, the evolution terms 
$e^{-i(t-s)H(\bm\phi)}$ and 
$e^{-isH(\bm\phi)}$ in step 3 
can be implemented by using 
the quantum gates $U(t-s,\bm\phi)$ 
and $U(s,\bm\phi)$, respectively. 
The remaining term $e^{-it\mu [\partial_{\phi_j}H(\bm\phi)]}$ 
in step 3 yields $e^{-it\mu H_j}$, 
which can be implemented by 
the quantum gate $U\big(t\mu,\bm e_j\big)$, 
where $\bm e_j$ is a unit vector with 1 
at the $j^{\rm th}$ element 
and zeros for the others. 
Therefore, all the evolution terms 
can be implemented by the device.
The density states $\rho^+$ and $\rho^-$ 
can be extracted and subtracted 
from each other using classical computers 
or quantum subtraction technology 
in real hardware,
e.g., see Ref.~\cite{doi:10.1142/S0219749919500564}. 

So far, the accuracy of an approach 
(such as finite difference, Stand.PSR, and Stoc.PSR)
is determined by its variance, which is a statistical error
raising from a finite number of measurements.
The variance of the Stoc. PSR is comparable 
with that of the Stand. PSR when 
an infinite number of measurements are taken. \cite{Banchi2021measuringanalytic}

\subsection*{Applications}
To demonstrate advantaged 
features of the Stoc.PSR method
for quantum metrology, 
we scrutinize the estimation in
two cases of single and multiple magnetic fields. 

\subsubsection*{\textbf {Single parameter estimation}}
\begin{figure}[t]
\centering
\includegraphics[width=8.6cm]{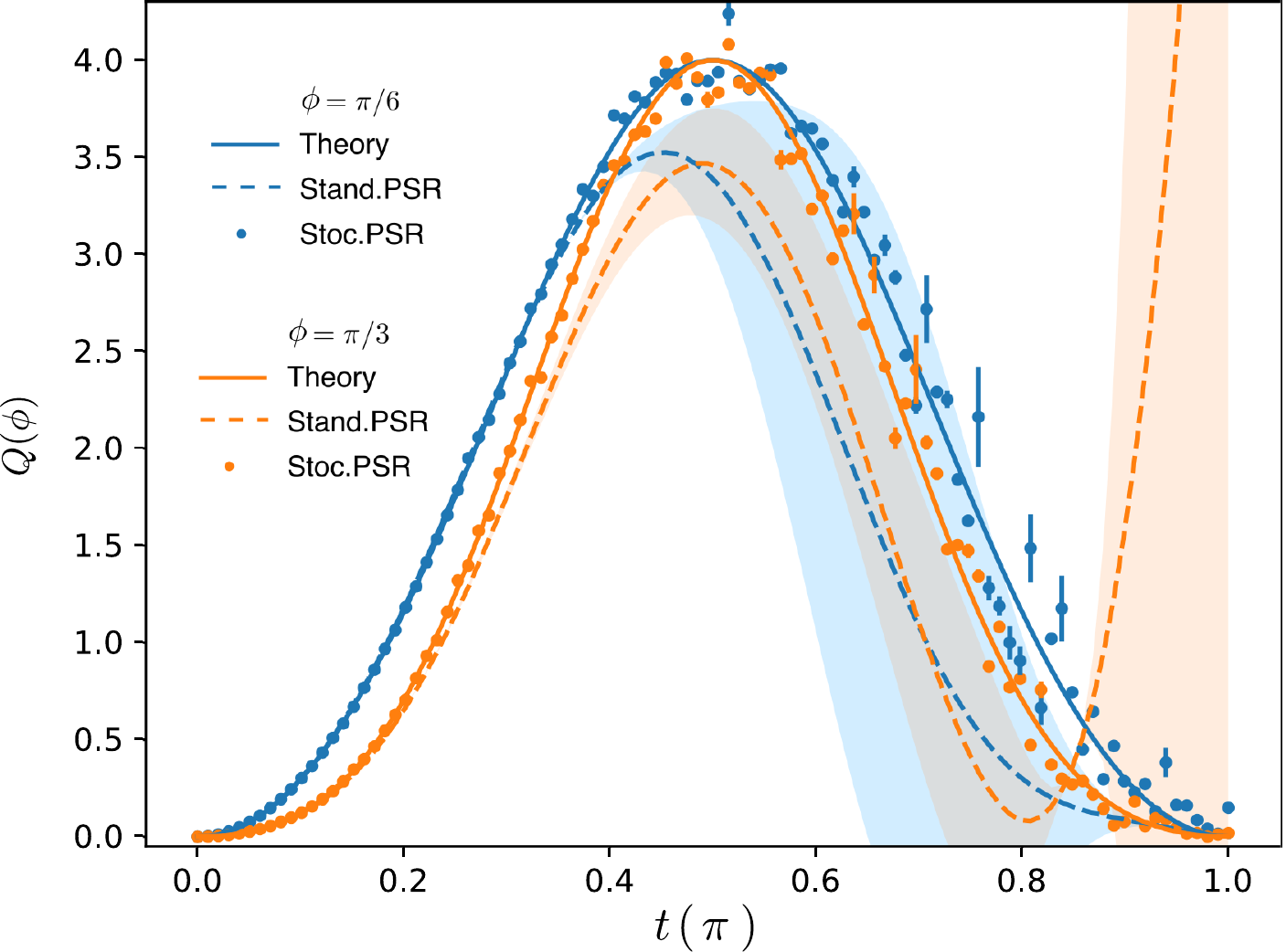}
\caption{\textbf{Quantum Fisher information for 
single magnetic field estimation.}
Quantum Fisher information $Q(\phi)$
as a function of the interaction time $t$
for different choices of $\phi$, as shown in
the figure. The solid curves are exact results
from theoretical analysis Eq.~\eqref{eq:theoF}, 
the dashed curves are obtained from 
the Trotter-Suzuki transformation and Stand.PSR,
and the dotted curves are obtained from the Stoc.PSR. 
It can be observed that $Q(\phi)$ varies 
with time $t$ and reaches 
its maximum at $t = \pi/2$.
More importantly, the results show 
that the Stoc.PSR agrees with the exact analysis
while the Stand.PSR gradually deviates from the exact one.
The mean-square error (MSE) 
are plotted as error bars and error areas in the figure.
They are systematic errors that caused by 
different calculation methods.
}
\label{fig:2}
\end{figure}

Let us consider a magnetic field 
$\bm B = \cos(\phi)\bm e_x + \sin(\phi)\bm e_z$,
and our goal is to estimate the angle $\phi$ between the field's
direction and the $z$ axis \cite{PhysRevA.90.022117}.
The field interacts with an exposed qubit probe and imprints its
information into the probe via the interaction Hamiltonian
\begin{align}\label{eq:Hi}
    H(\phi) = \bm B\cdot\bm\sigma = 
    \cos(\phi)\sigma_x + \sin(\phi)\sigma_z,
\end{align}
where $\bm \sigma = (\sigma_x, \sigma_y, \sigma_z)$
are the Pauli matrices. The unitary evolution is given by
$U(t,\phi) = e^{-itH(\phi)}$.
Applying this transformation, 
an initial probe state, i.e., 
$|\psi_0\rangle = \big(|0\rangle 
+ |1\rangle\big)/\sqrt{2}$
evolves to $|\psi(\phi)\rangle = U(t,\phi)|\psi_0\rangle$.
The evolved probe state $|\psi(\phi)\rangle$
provides the best quantum strategy for the 
estimation of $\phi$, which can be evaluated 
via the QFI, similar to
Eq.~\eqref{eq:Qpure1}
\begin{align}\label{eq:theoF}
\notag  Q(\phi) &=4{\rm Re}
    \big[
    \langle\psi_0|Y_\phi^2|\psi_0\rangle
    -|\langle\psi_0|Y_\phi|\psi_0\rangle|^2
    \big]\\
   & = 4\sin^2(t)
   \big[1-\cos^2(t)\sin^2(\phi)\big],
\end{align}
where 
$
    Y_\phi = 
     \int_0^t
    e^{isH(\phi)}\big[\partial_{\phi} 
    H(\phi)\big]e^{-isH(\phi)}
    {\rm d}s
$ 
(see detailed in the Methods section).
The QFI $Q(\phi)$ 
is time-dependent and 
achieves a maximum 
value of 4 at $t = \pi/2$,
as shown by  
the solid curves in Fig.~\ref{fig:2}.
This behavior 
is caused 
by  the rotation 
of the probe state
under 
magnetic field.
Furthermore, the QFI depends on the 
true parameter value, it thus
becomes a function of $\phi$.
In the limit $\phi\to 0$, the QFI yields 
$Q(\phi) = Q_{\rm max} = 4\sin^2(t)$
\cite{PhysRevA.90.022117}.

We now apply the Stoc.PSR 
to a single-qubit quantum circuit.
The circuit is initially 
prepared in $|0\rangle$, 
and it becomes $|\psi_0\rangle$ after 
applying a Hadamard gate. 
Using 
the definition $\partial_\phi |\psi(\phi)\rangle
= \big[\partial_\phi U(t,\phi)\big]|\psi_0\rangle$,
and the first line in Eq.~\eqref{eq:dU},
we have
\begin{align}\label{eq:stoc_psi}
\dfrac{\partial|\psi(\phi)\rangle}
{\partial\phi} = 
-i \int_0^t U(t,\phi) O_\phi |\psi_0\rangle\
 {\rm d}s,
\end{align}
where 
$O_\phi = e^{isH(\phi)}[\partial_\phi H(\phi)]
e^{-isH(\phi)}$.
Similar as above, we have
\begin{align}\label{eq:Opsi0}
O_\phi|\psi_0\rangle
= \dfrac{i}{2\sin(\mu t)}
\Big[e^{-it\mu O_\phi}-e^{it\mu O_\phi}\Big]
|\psi_0\rangle,
\end{align}
where $t\mu \notin \pi \mathbb{Z}$.
Using $e^{-it\mu O_\phi}
= e^{isH(\phi)}e^{-it\mu [\partial_\phi H(\phi)]}
e^{-isH(\phi)}$, we derive Eq.~\eqref{eq:stoc_psi}
as
\begin{align}\label{eq:stoc_psi1}
\dfrac{\partial|\psi(\phi)\rangle}
{\partial\phi} 
 = \dfrac{1}{2\sin(t\mu)}
    \int_0^t\Big[|\psi^+\rangle
    - |\psi^-\rangle\Big] {\rm d}s,
\end{align}
where
$|\psi^\pm\rangle$ are
given by 
\begin{align}\label{eq:psipm}
    |\psi^\pm\rangle = U(t-s,\phi)\cdot
    e^{\mp it\mu[\partial_\phi H(\phi)]}\cdot
    U(s,\phi)|\psi_0\rangle.
\end{align}
In the numerical calculation,
we derive 
$\partial_\phi|\psi(\phi)\rangle
=\frac{t}{N*2\sin(t\mu)}
\sum_{n=1}^N [|\psi^+\rangle -
|\psi^-\rangle]$
with $N$ samplings of
$s\in [0,t]$.
This is a simplified version of 
Algorithm \ref{al:1} for pure states.
We set $N = 1000$ 
and obtain the QFI $Q(\phi)$ which is of the form
\eqref{eq:Qpure}
\begin{align}\label{eq:QphiStoc}
\hspace{-0.5cm}
    Q(\phi) = 
    \dfrac{t^2}{N^2\sin^2(t\mu)}{\rm Re}\Big[\langle\Psi|\Psi\rangle
    -\big|\langle \Psi|\psi(\phi)\rangle\big|^2\Big],
\end{align}
where $|\Psi\rangle = 
\sum_{n=1}^N\big[|\psi^+\rangle-|\psi^-\rangle\big]$.

To implement the Stoc.PSR in quantum computers, 
we assume there exists 
a universal quantum hardware
that allows for executing
the quantum gate $U(t,\phi)$. 
Changing the variables in $U(t,\phi)$
by $U(x,z) = e^{-it(x\sigma_x + z\sigma_z)}$
where $x = \cos(\phi)$ and $z = \sin(\phi)$,
it yields $\partial_\phi U(x,z) = 
\partial_x U(x,z)\partial_\phi x
+\partial_z U(x,z)\partial_\phi z$.
This is a universal quantum device 
because all the evolution terms in 
Eq.~\eqref{eq:psipm} can be implemented 
via this quantum gate in the device.

Finally, let us compare the results with 
the Stand.PSR. 
To apply the Stand.PSR,
we first decompose the evolution $U(t,\phi)$
into a sequence of sub-evolutions 
through Trotter-Suzuki transformation
\cite{Dhand_2014}
\begin{align}\label{eq:Utrotter}
    U(t,\phi) = \lim_{m\to\infty}
   \big(e^{-it\cos(\phi)\sigma_x/m}
    e^{-it\sin(\phi)\sigma_z/m}
    \big)^m,
\end{align}
where these sub-evolutions 
can be executed in quantum circuits 
through rotation gates, specifically $Rx$ and $Rz$. 
The derivative $\partial_\phi|\psi(\phi)\rangle$
now can be implemented by using
the Stand.PSR. 
See detailed calculation in the Method section.


Figure~\ref{fig:2} shows a comparison between the 
performance of Stand.PSR and Stoc.PSR
with the exact theoretical result. 
The Stoc.PSR consistently demonstrates
a good agreement with the 
exact results all the time
while the Stand.PSR deviates from the 
exact results as time increases.
It implies that using Stoc.PSR 
in quantum circuits for studying 
quantum systems with generic Hamiltonian 
is essential and cannot be replaced 
by similar approximation methods.
This is further supported by 
considering the mean-square error 
(MSE), defined as $(1/M)\sum_i [y_i(t) - f(t)]^2$, 
where $M$ denotes the number of data points,
$y_i(t)$ represents the data 
obtained using the Stand.PSR or 
Stoc.PSR and $f(t)$ represents 
the exact results given by Eq.~\eqref{eq:theoF}. 
We emphasize that the MSE here plays no role
with the error of the estimated parameter,
it is rather a systematic error caused by different 
methods when comparing 
with the exact theoretical result.
The MSEs are shown in the figure as the error bars
and error areas.
As we can see,
the MSE for Stoc.PSR remains 
small throughout the duration, 
while 
that one for the 
Stand.PSR divers for large sensing time $t$.

\subsubsection*{Multiple parameters estimation}
Next, we apply the Stoc.PSR 
scheme to estimate the components 
of a magnetic field 
pointing in an arbitrary direction. 
Consider the probe state
initially prepared in $n$-qubit GHZ state
$|\psi_0\rangle = \big(|00\cdots0\rangle 
+ |11\cdots1\rangle\big)/\sqrt{2}$,
such that allows for obtaining the maximum QFIM
\cite{PhysRevLett.96.010401}.
The interaction Hamiltonian is given by
\begin{align}\label{eq:Hmul}
    H(\bm\phi) = \sum_j\phi_jJ_j,\quad 
    \text{for } j\in\{x, y, z\},
\end{align}
where $\bm\phi 
= (\phi_x, \phi_y, \phi_z)$ are three components
of the given magnetic field that we want to estimate,
and $J_j = \sum_{k=1}^n\sigma_j^{(k)}$ is 
a collective Pauli matrix. 
Potential platforms for the probe
include spin-1/2 ensemble semiconductors, 
ions traps, NMR systems, and NV centers.
In these systems, such as
spin-1/2 ensemble, $J_j$ becomes 
the collective angular momentum operator
\cite{PhysRevA.102.022602}.

\begin{figure}[t]
\centering
\includegraphics[width=8.6cm]{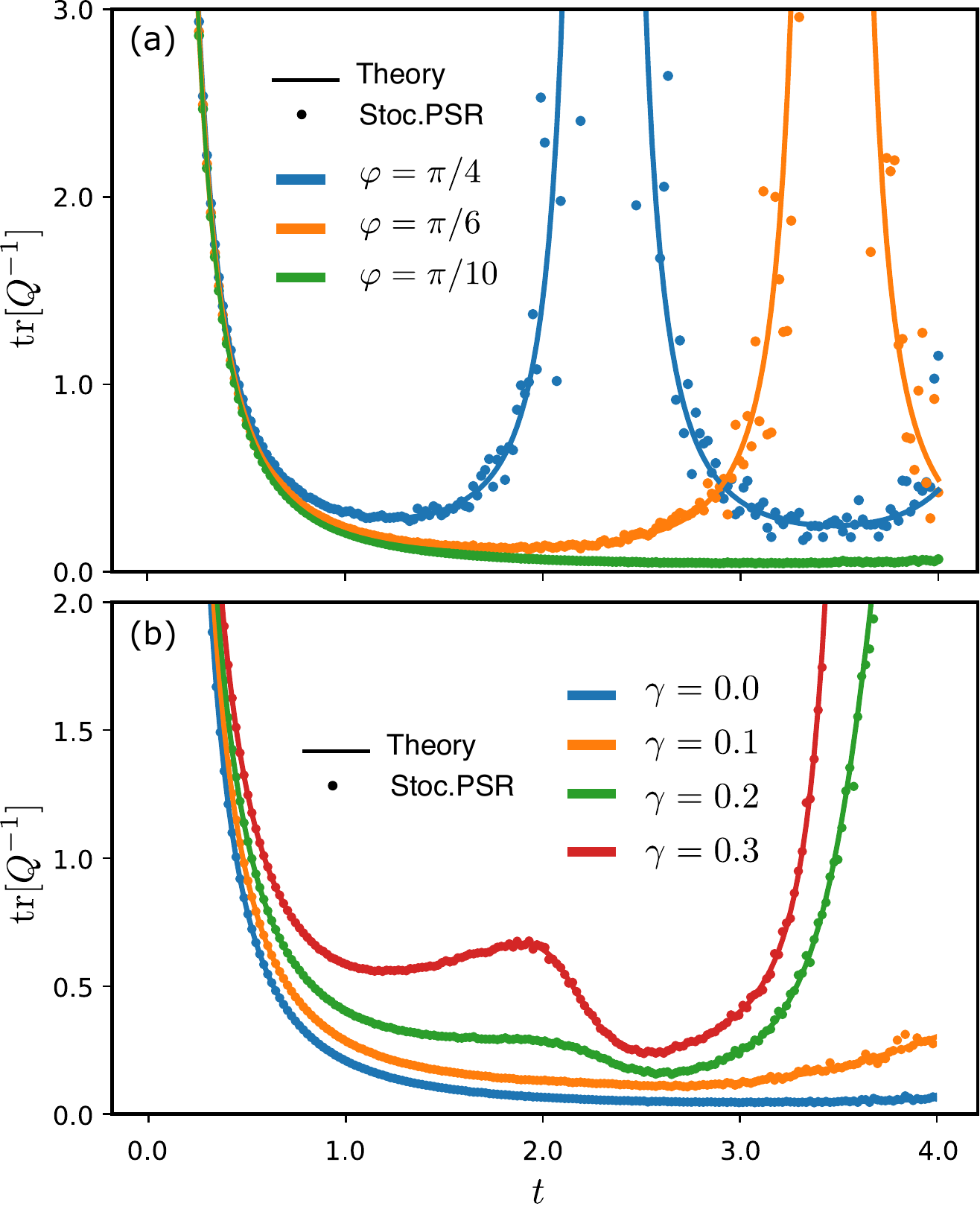}
\caption{\textbf{
The total variance for
multiphase magnetic field estimation.}
(a) The total variance
 $\Delta^2\bm\phi = {\rm tr}[Q^{-1}]$
as a function of the interaction time $t$
for different choices of $\varphi$ as shown in
the figure. Here, we use $ 
\phi_x = \phi_y = \phi_z = \varphi$
for 
illustration
(although in general, these values may differ).
The solid curves are exact results, which are given
by theoretical analysis,
and the dotted curves are obtained from the Stoc.PSR. 
It can be observed that ${\rm tr}[Q^{-1}]$ varies 
with $t$ and reaches 
its minimum at a certain time.
More importantly, the results show 
a good agreement between 
the Stoc.PSR  
and the exact theoretical analysis.
(b) Plot of the total variance versus $t$
under the time-dependent dephasing noise
for various decay rates $\gamma$.
Here, we fixed $\varphi = \pi/10$.
}
\label{fig:3}
\end{figure}

The QFIM can be 
obtained theoretically from Eq.~\eqref{eq:Qpure1},
and the total variance yields
$\Delta^2\bm\phi = {\rm tr}[Q^{-1}]$.
Concretely, with $n = 3$ qubits 
and $\phi_x = \phi_y = \phi_z = \varphi$,
we obtain
\begin{align}
    {\rm tr}[Q^{-1}] = \dfrac{7}{108t^2}
    +\dfrac{3\varphi^2}{54\sin^2(\sqrt{3}\varphi t)}.
\end{align}
We show the exact theoretical results by 
the solid curves for various $\varphi$
in Fig.~\ref{fig:3}a.
For each $\varphi$,
there is a minimum variance 
at a certain time $t$,
which is caused by 
the rotation of the probe state
under 
magnetic field. 
In the limit of small phase, i.e., 
$\varphi\to 0$, the total variance is 
 ${\rm tr}[Q^{-1}] = \frac{7}{108t^2}$,
which results in the minimum of total variance. 

In the Stoc.PSR method, 
we model the probe in an $n$-qubit quantum circuit 
initially prepared in 
the GHZ state. 
The circuit can be implemented
in the existing noisy intermediate-scale 
quantum computers 
\cite{Preskill2018quantumcomputingin}.
Its state evolves under the transformation 
$U(t,\bm\phi)= e^{-itH(\bm\phi)}$, 
and results in the evolved 
state $|\psi({\bm\phi})\rangle
= U(t,\bm\phi)|\psi_0\rangle$.
As discussed above, this 
unitary evolution
can be implemented 
in a universal quantum computer.
Therefore, we employ the Stoc.PSR
using Algorithm~\ref{al:1} to obtain 
$\partial_{\phi_j}|\psi(\bm\phi)\rangle$ for all $j$ 
and get the QFIM as in Eq.~\eqref{eq:Qpure}.
The ${\rm tr}[Q^{-1}]$ is shown in Fig.~\ref{fig:3}a
(dotted curves) 
for the number of sampling $N = 1000$.
The Stoc.PSR's results
agree with the exact
results.

We further apply the scheme to noisy cases,
where the probe is described by mixed states.
We consider time-dependent 
dephasing, 
which is given by a quantum channel $\mathcal{E}$
that acts on a single qubit 
as
\begin{align}\label{eq:rhoty}
\mathcal{E} [\rho] := K_1\rho K_1^\dagger 
+ K_2\rho K_2^\dagger,
\end{align}
where 
we used the Kraus representation
for the dephasing channel \cite{Koczor_2020}
\begin{align}\label{eq:kraus}
K_1 = \begin{pmatrix}
	p(t) & 0 \\
	0 & 1
	\end{pmatrix}, \
K_2 = \begin{pmatrix}
	\sqrt{1-p^2(t)} & 0 \\
	0 & 0
	\end{pmatrix}.
\end{align}
The time-dependent probability is
$p(t) = e^{-\gamma t}$ for the Markovian noise,
where $\gamma$ is the decay rate \cite{Koczor_2020}.

We apply the quantum channel $\mathcal{E}$
to all qubits in the probe during the interaction time
and use
Algorithm \ref{al:1} to derive the QFIM.
The results for the total variance versus the interaction time $t$
are shown in Fig.~\ref{fig:3}b.
We plot the results for several decay rates $\gamma$ 
and compare the Stoc.PSR approach 
with the theoretical analysis. 
Again, they match excellently. 

\section*{Discussion}

We additionally discuss the application
to Hamiltonian tomography in many-body systems,
which involves determining 
unknown coupling constants in the Hamiltonian.
Hamiltonian
tomography
aims to reconstruct a generic many-body Hamiltonian
by measuring multiple pairs of the initial 
and time-evolving states.
It is a challenging task due to the complexity of 
the many-body dynamics. So far, the progress is limited 
to particular Hamiltonians and small-size systems 
\cite{PhysRevA.95.062120,
Wang_2015,PhysRevLett.112.190501,
PhysRevLett.102.187203}.
For example, a simple task is to identify 
the Hamiltonian in an 
Ising model of a spin-1/2 chain placed 
under an external field. 
A generic Hamiltonian is
given by $H = 
\sum_j c_{j,j+1}\sigma_z^{(j)}
\sigma_z^{(j+1)} + 
\sum_j h_j
\sigma_x^{(j)} $,
where the coupling constants $\{c_{j,j+1}\}$ 
and the external field strengths $\{h_j\}$
are unknown factors,
$j$ stands for the site $j^{\rm th}$
in the chain.

Recently, Li et al. introduced a quantum quench 
approach for the Hamiltonian tomography 
that can apply to both analog and digital 
quantum simulators \cite{PhysRevLett.124.160502}.
Hereafter, we evaluate the quantum quench 
precision by using Stoc.PSR 
to calculate the classical Cram{\'e}r-Rao bound.

A generic Hamiltonian of 
a many-body system can be decomposed 
into $d$-interaction terms as
\begin{align}\label{eq:Hmany}
H = \sum_{j = 1}^d x_j H_j,
\end{align}
where $\{x_j\}$ are unknown coupling constants
that need to be determined,
and $\{H_j\}$ are Hermitian operators.
An initial state $\rho_0$
evolves to $\rho(\bm x)
= e^{-iHt}\rho_0e^{iHt}$
after time $t$,
for $\bm x = \big(x_1, \cdots, x_d\big)^\intercal$.
The system obeys a conservation law
\cite{PhysRevLett.124.160502}
\begin{align}\label{eq:convlaw}
{\rm tr}\big[\rho_0H\big] = 
{\rm tr}\big[\rho(\bm x)H\big],
\end{align}
for every pair of given $\rho_0$ and  $\rho(\bm x)$.
To determine $d$ coefficients $\{x_j\}$,
we need to solve at least 
$p \ge d-1$ linear equations
which form a matrix equation as 
${\bm X} \bm x = \bm 0$, where
${\bm X}$  is a $p\times d$ 
matrix with the elements
\begin{align}\label{eq:Mmax}
{X}_{k,l} = 
{\rm tr}\big[\rho_0^{(k)}H_l\big] -
{\rm tr}\big[\rho^{(k)}(\bm x)H_l\big],
\end{align}
where $k\in\{1,\cdots, p\}$ and 
$l\in\{1, \cdots, d\}$ for different pairs of
$\rho_0^{(k)}, \rho^{(k)}(\bm x)$.
Here, $\big\{\rho_0^{(k)}\big\}$
is a set of (random) initial states 
and $\big\{\rho^{(k)}(\bm x)\big\}$
is a set of evolved states.

For $\{H_j\}$ are measured observables, 
such as Pauli matrices, 
SIC-POVM, and polarization bases
\cite{PhysRevA.104.052431},
the matrix elements $\{X_{k,l}\}$ 
become measured probabilities under 
the eigenbases of these observables.
Thus, to evaluate the best estimation of $\{x_j\}$,
we examine the classical bound, 
i.e., via the CFIM  
Eq.~\eqref{eq:F}. Firstly, 
from Eq.~\eqref{eq:Mmax}, we derive 
\begin{align}\label{eq:deri_X}
\dfrac{\partial X_{k,l}}{\partial x_j}
= - {\rm tr}\bigg[
\bigg(\frac{\partial\rho^{(k)}(\bm x)}
{\partial x_j}\bigg)H_l
\bigg],
\end{align}
where $\frac{\partial\rho^{(k)}(\bm x)}
{\partial x_j}$ is given by Stoc.PSR 
Eq.~\eqref{eq:drhoFi}.
We later define the CFIM as
\begin{align}\label{eq:F_TM}
    F_{i,j} = \sum_{\{k,l\}} \dfrac{1}{|X_{k,l}|}
   \big[\partial_{x_i} X_{k,l}\big]
   \big[\partial_{x_j} X_{k,l}\big],
   \forall i,j \in \{1,\cdots, d\},
\end{align}
and hence 
obtain the classical Cram{\'e}r-Rao bound,
i.e., $\Delta^2\bm x \ge {\rm tr}[F^{-1}]$.
The equality can be achieved 
by an appropriate estimator.  

For numerical demonstration, we consider 
a single-qubit system
whose Hamiltonian is given 
by 
\cite{PhysRevA.104.052431}
\begin{align}\label{eq:ising2}
H = \sum_{i = 1}^3 x_i|\psi_i\rangle\langle\psi_i|,
\end{align}
where $\{x_i\}$ are unknown coefficients,
$|\psi_1\rangle = |0\rangle,
|\psi_2\rangle = \big(|0\rangle + |1\rangle\big)/\sqrt{2}$,
and
$|\psi_3\rangle = \big(|0\rangle + i|1\rangle\big)/\sqrt{2}$.
We apply the quantum quench method 
to find $\{x_i\}$
and analyze the variance $\Delta^2\bm x$.
It is given by the classical Cram{\'e}r-Rao bound,
i.e., $\Delta^2\bm x = {\rm tr}[F^{-1}]$.
The derivative $\frac{\partial\rho^{(k)}(\bm x)}
{\partial x_j}$ in Eq.~\eqref{eq:deri_X} 
is given by either Stoc.PSR 
or finite difference approach for comparison. 
For the Stoc.PSR, we run 1000 random 
samples of $s \in [0,1], t = 1, \mu = \pi/4$.
For the finite difference method,
we use $\partial_x\rho(x) = \frac{
\rho(x+\epsilon) - 
\rho(x-\epsilon)}{2\epsilon}$,
where $\epsilon$ is a small step size.
The variance $\Delta^2\bm x$ is averaged 
after 10 repetitions.

\begin{figure}[t]
\centering
\includegraphics[width=8.6cm]{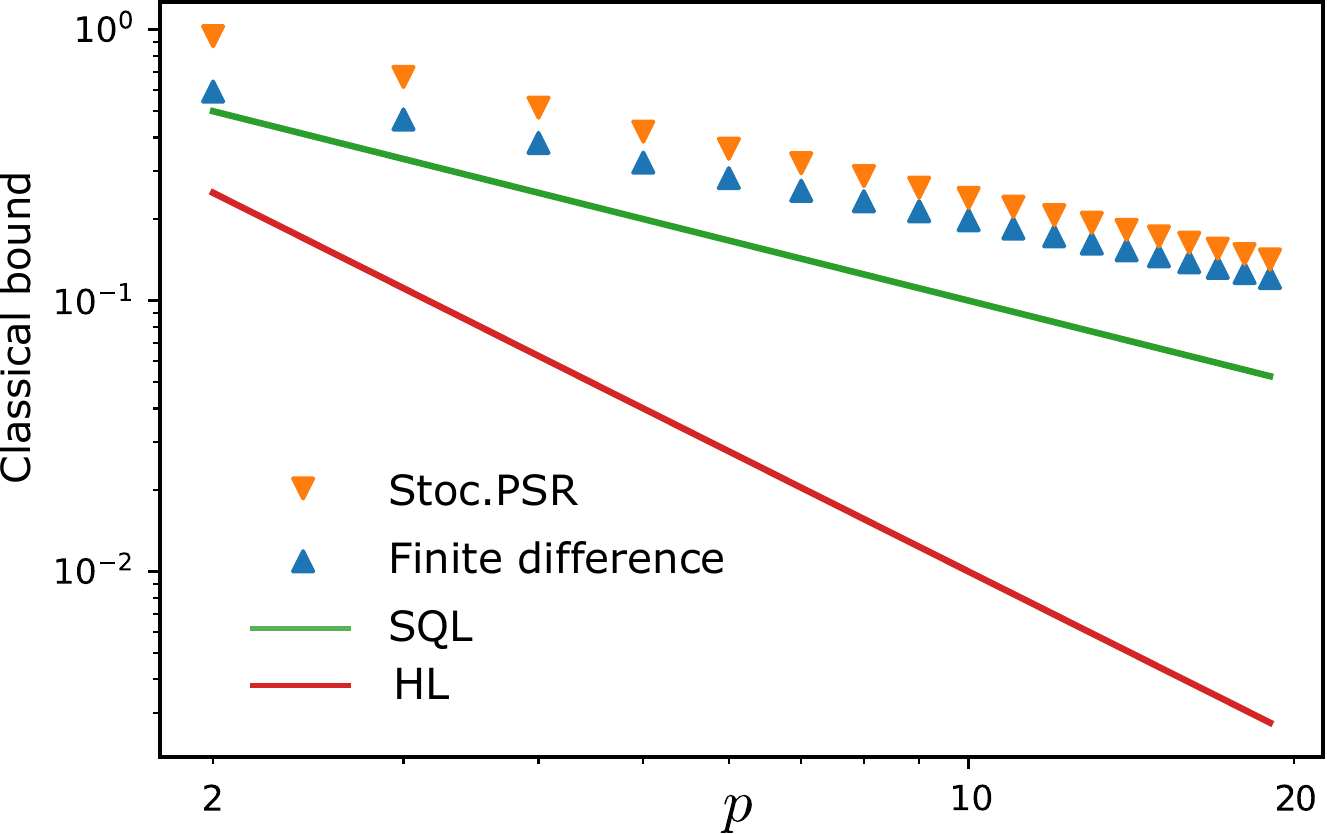}
\caption{\textbf{
The classical Cram{\'e}r-Rao bound
in single-qubit Hamiltonian tomography.}
The results are plotted for two different approaches:
Stoc.PSR (down triangle) and finite difference (up triangle).
The standard quantum limit (SQL) and 
Heisenberg limit (HL) are shown for comparison. 
Here, SQL $\propto 1/p$ and HL $\propto 1/p^2$.}
\label{fig:4}
\end{figure}

The results are shown in Fig.~\ref{fig:4}
as functions of $p$.
In principle, $p = d-1$
is sufficient to estimate $d$ parameters.
However, $p > d-1$ gives better statistical results
\cite{PhysRevLett.124.160502}.
In this context, we compare the Stoc.PSR and  
finite difference approach,
and find that they tend to converge 
when increasing $p$.
While the finite difference method 
consistently produces better results, 
it also has a larger bias due 
to the computational challenges 
of computing $\rho(x+\epsilon)$ and 
$\rho(x-\epsilon)$ in quantum circuits for $\epsilon \ll 1$.
Furthermore, since $p$ is equivalent with the number of
repeated measurements,
we can define the standard quantum limit
(SQL) as $\propto 1/p$ and Heisenberg limit (HL) by  $\propto 
1/p^2$. 
We compare the bound in quantum quench
with these limits and find that 
 it scales slightly worse than the SQL,
 opening further exploring to improve
 the limit in quantum quench approaches. 

\section*{Methods}
\subsection*{Proof of time-dependent 
stochastic parameter-shift rule}\label{appA}
We consider the derivative of a mixed quantum state 
as in Eq.~\eqref{eq:drho_re} in the main text
\begin{align}\label{eq:drho_re_app}
    \dfrac{\partial \rho(\bm\phi)}
    {\partial\phi_j} = -i\int_0^t U(\bm\phi)
    \big[O_j,\rho_0\big]U^\dagger(\bm\phi) ds,
\end{align}
where $ O_j = e^{isH(\bm\phi)}\big[\partial_{\phi_j} 
    H(\bm\phi)\big]e^{-isH(\bm\phi)}
$. We have
\begin{align}\label{eq:dZ_app}
    \big[O_j,\rho_0\big]
    = \dfrac{i}{\sin(2t\mu)}
    \Big[e^{-it\mu O_j}\rho_0e^{it\mu O_j} - 
    e^{it\mu O_j}\rho_0e^{-it\mu O_j}\Big],
\end{align}
\textbf{Proof:}
Using the Baker-Campbell-Hausdorff formula 
\cite{Achilles2012},
we derive
\begin{align}\label{eq:BCH_app}
\notag e^{-it\mu O_j}\rho_0e^{it\mu O_j} 
= \rho_0 &+ \big[-it\mu O_j, \rho_0\big] \\
& + \dfrac{1}{2!}
\Big[-it\mu O_j,\big[-it\mu O_j,\rho_0\big]\Big] + \cdots
\end{align}
and 
\begin{align}\label{eq:BCH_app1}
\notag e^{it\mu O_j}\rho_0e^{-it\mu O_j} 
= \rho_0 &+ \big[it\mu O_j, \rho_0\big] \\
& + \dfrac{1}{2!}
\Big[it\mu O_j,\big[it\mu O_j,\rho_0\big]\Big] + \cdots
\end{align}
Subtracting Eq.~\eqref{eq:BCH_app1} from
Eq.~\eqref{eq:BCH_app} yields
\begin{align}\label{eq:umd_app}
\notag &\Big[e^{-it\mu O_j}\rho_0e^{it\mu O_j} - 
    e^{it\mu O_j}\rho_0e^{-it\mu O_j}\Big] = \\
&-2i\dfrac{t\mu}{1!}\big[O_j,\rho_0\big]
+2i\dfrac{(t\mu)^3}{3!}
\bigg[O_j,\Big[O_j,\big[O_j,\rho_0\big]\Big]\bigg] 
-2i\dfrac{(t\mu)^5}{5!}\cdots
\end{align}
where using the algebraic expansion with
the condition $O_j^2 = \bm I$, we have
$\bigg[O_j,\Big[O_j,\big[O_j,\rho_0\big]\Big]\bigg] 
= \dfrac{2^3}{2}\big[O_j,\rho_0\big]$,
and so on. 
Finally, Eq.~\eqref{eq:umd_app} becomes
\begin{align}\label{eq:thl_app}
\Big[e^{-it\mu O_j}\rho_0e^{it\mu O_j} - 
    e^{it\mu O_j}\rho_0e^{-it\mu O_j}\Big]
    = -i\sin(2t\mu)\big[O_j,\rho_0\big].
\end{align}
Multiplying two sides of Eq.~\eqref{eq:thl_app} 
by $\frac{i}{\sin(2t\mu)}$ we arrive at Eq.~\eqref{eq:dZ_app} 

Now, substituting Eq.~\eqref{eq:dZ_app}
into Eq.~\eqref{eq:drho_re_app}, we have
\begin{align}\label{eq:drho_re1_app}
\notag    \dfrac{\partial \rho(\bm\phi)}
    {\partial\phi_j} = \dfrac{1}{\sin(2t\mu)}
    & \int_0^t U(\bm\phi)
    \Big[e^{-it\mu O_j}\rho_0e^{it\mu O_j} - \\
    & e^{it\mu O_j}\rho_0e^{-it\mu O_j}\Big]
    U^\dagger(\bm\phi) ds.
\end{align}
Using $ e^{-it\mu O_j} = e^{isH(\bm\phi)}
e^{-it\mu \big[\partial_{\phi_j} 
H(\bm\phi)\big]}
e^{-isH(\bm\phi)}
$ and $U(\bm\phi) = e^{-itH(\bm\phi)}$,
we set
\begin{align}\label{eq:ue1_app}
\notag U_j^\pm(\bm\phi,s) &= U(\bm\phi)e^{\mp it\mu O_j} \\
\notag &= e^{-itH(\bm\phi)}
e^{isH(\bm\phi)}
e^{\mp it\mu \big[\partial_{\phi_j} H(\bm\phi)\big]}
e^{-isH(\bm\phi)}\\
& = e^{-i(t-s)H(\bm\phi)}
e^{\mp it\mu \big[\partial_{\phi_j} H(\bm\phi)\big]}
e^{-isH(\bm\phi)}.
\end{align}
Substituting Eq.~\eqref{eq:ue1_app}
into Eq.~\eqref{eq:drho_re1_app}, we obtain
\begin{align}\label{eq:drho_re2_app}
\notag    \dfrac{\partial \rho(\bm\phi)}
    {\partial\phi_j} &= \dfrac{1}{\sin(2t\mu)}
    \int_0^t \bigg[U_j^+(\bm\phi,s)
    \rho_0\big[U_j^+(\bm\phi,s)\big]^\dag \\  
\notag    &\hspace{3cm} -U_j^-(\bm\phi,s)
    \rho_0\big[U_j^-(\bm\phi,s)\big]^\dag\bigg] ds \\
    & = 
    \dfrac{1}{\sin(2t\mu)}
    \int_0^t \bigg[ \rho_j^+(\bm\phi,s)
    - 
    \rho_j^-(\bm\phi,s)\bigg]
    ds,
\end{align}
where we used
$\rho_j^\pm(\bm\phi,s)
= U_j^\pm(\bm\phi,s)\rho_0 \big[U_j^\pm(\bm\phi)\big]^\dagger.
$

\subsection*{Theoretical analysis for 
single-parameter estimation}\label{appB}
Firstly, let us discuss the exact calculation method 
for quantum Fisher information in 
single parameter estimation.
Starting from 
$H(\phi) = \cos(\phi)\sigma_x 
+ \sin(\phi)\sigma_z$, we derive 
$\partial_\phi H(\phi) = -\sin(\phi)\sigma_x +
\cos(\phi)\sigma_z$.
Substituting it into $Y_j$ for $j = \phi$, 
we obtain
\begin{align}
\hspace{-0.75cm}
\notag    Y_\phi &= 
     \int_0^t
    e^{isH(\phi)}\big[\partial_{\phi} 
    H(\phi)\big]e^{-isH(\phi)}
    {\rm d}s \\
\notag    & = \dfrac{1}{2}
    \begin{pmatrix}
   \sin2t\cos\phi & -\sin2t\sin\phi - 2i\sin^2t\\
   -\sin2t\sin\phi + 2i\sin^2t & -\sin2t\cos\phi
    \end{pmatrix}.
\end{align}
Finally, we derive the 
quantum Fisher information
as in Eq.~\eqref{eq:Qpure}:
\begin{align}\label{eq:Qphipure}
\notag    Q(\phi) &=4{\rm Re}
    \big[
    \langle\psi_0|Y_\phi^2|\psi_0\rangle
    -|\langle\psi_0|Y_\phi|\psi_0\rangle|^2
    \big]\\
    & = 4\sin^2(t)
   \big[1-\cos^2(t)\sin^2(\phi)\big],
\end{align}
which results in Eq.~\eqref{eq:theoF}.

\setcounter{equation}{0}
\renewcommand{\theequation}{C.\arabic{equation}}
\subsection*{Trotter-Suzuki transformation and Stand.PSR}\label{appC}
From now on, let us show the detailed calculation for 
the Trotter-Suzuki transformation and Stand.PSR 
for single parameter estimation. 
From the evolution~\eqref{eq:Utrotter}, we set
\begin{align}\label{eq:xz}
    \begin{cases}
      x = \frac{2t}{m}\cos(\phi) \\
      z = \frac{2t}{m}\sin(\phi)
    \end{cases}
    \Rightarrow 
     \begin{cases}
      \partial_\phi x = -\frac{2t}{m}\sin(\phi) \\
      \partial_\phi z = \frac{2t}{m}\cos(\phi)
    \end{cases}.
\end{align}
Then, Eq.~\eqref{eq:Utrotter} is recast as
\begin{align}\label{eq:Urecase}
    U(x,z) = 
   \lim_{m\to\infty}\big(e^{-i\frac{x}{2}\sigma_x}
   e^{-i\frac{z}{2}\sigma_z}\big)^m,
\end{align}
and thus
\begin{align}\label{eq:dUrecase}
    \partial_\phi U(x,z) = 
    \dfrac{\partial U(x,z)}{\partial x}
    \dfrac{\partial x}{\partial \phi} +
    \dfrac{\partial U(x,z)}{\partial z}
    \dfrac{\partial z}{\partial \phi}.
\end{align}
Concretely, we have
\begin{align}
    \partial_x U(x,z) &= \dfrac{m}{2}(-i\sigma_x)U(x,z), \label{eq:dUx0}\\
    \partial_z U(x,z) &= \dfrac{m}{2}(-i\sigma_z)U(x,z). \label{eq:dUz0}
\end{align}
Note that $U(\pi,0) = \underset{m\to\infty}{\lim}(-i\sigma_x)^m$.
For $m = 4k+1\ \forall k\in \mathbb{N}$,
we have $U(\pi,0) = -i\sigma_x$, 
from which the Pauli matrix 
$\sigma_x$ can be implemented 
by the unitary (quantum gate) $U(\pi,0)$.
Likewise, $U(0,z+\pi) = -i\sigma_z$.
Now, Eqs. (\ref{eq:dUx0}, \ref{eq:dUz0}) are recast as
\begin{align}
    \partial_x U(x,z) &
    = \dfrac{m}{2}U(x+\pi,z), \label{eq:dUx}\\
    \partial_z U(x,z) &
    = \dfrac{m}{2}U(x,z+\pi). \label{eq:dUz}
\end{align}
Here, $m$ obeys the periodic property, 
therefore its choice will not affect the results.
Hence, these derivatives (\ref{eq:dUx}, \ref{eq:dUz})
can be obtained in quantum circuits 
by modifying the Stand.PRS.
Substituting  Eqs. (\ref{eq:dUx}, \ref{eq:dUz})
and Eq.~\eqref{eq:xz} into Eq.~\eqref{eq:dUrecase},
we derive
\begin{align}
\notag    \partial_\phi|\psi(\phi)\rangle
    &=\partial_\phi U(x,z)|\psi_0\rangle\\
   &=t\big[-\sin(\phi)U(x+\pi,z)
    +\cos(\phi)U(x,z+\pi)\big]
    |\psi_0\rangle,
\end{align}
where $|\psi_0\rangle$ is the initial probe state.
In this form, the QFI is given as
\begin{align}\label{eq:QbStand}
\hspace{-0.5cm}
   Q(\phi) &= 4{\rm Re}
    \big[\langle\partial_\phi\psi(\phi)|\partial_\phi\psi(\phi)\rangle
    -|\langle\partial_\phi\psi(\phi)|\psi(\phi)\rangle|^2
    \big].
\end{align}
The procedure for calculating 
the quantum Fisher information is shown
in Algorithm~\ref{al:2} below.

\begin{algorithm}
\caption{Standard parameter-shift rule}\label{al:2}
\KwData{$|\psi_0\rangle,\phi, U(x,z)$}
\KwResult{$Q(\phi)$}
$T \gets $ time (array) \;
$m \gets 4k+1\ \forall k \in \mathbb{N}$ \;
\For{t in T}{
  $x = 2t\cos(\phi)/m$\\
  $z = 2t\sin(\phi)/m$\\
  $dx = U(x+\pi,z)|\psi_0\rangle$\\
  $dz = U(x,z+\pi)|\psi_0\rangle$\\
  $d\psi = t[-\sin(\phi)dx + \cos(\phi)dz]$\\
  get $Q(\phi)$ from Eq.~\eqref{eq:QbStand}.
}
\end{algorithm}
%


\subsection*{Multiple parameters estimation}
Hereafter, we derive the multiple parameters estimation.
For $n = 3$, we first calculate $J_j$ for $j = \{x, y, z\}$
as 
\begin{align}
    J_j = \sigma_j\otimes I_2 \otimes I_2 + 
    I_2 \otimes \sigma_j\otimes I_2 + 
     I_2 \otimes I_2 \otimes \sigma_j,
\end{align}
where $I_2$ is the $2\times2$ identity matrix.
The Hamiltonian $H(\bm\phi)$ is given by Eq.~\eqref{eq:Hmul},
and its derivative yields 
$\partial_{\phi_j}H(\bm\phi)
= J_j$.
Similar to the above, we derive $Y_j$ 
\begin{align}\label{eq:Ybmphi}
    Y_j &= 
     \int_0^t
    e^{isH(\bm\phi)}J_je^{-isH(\bm\phi)}
    {\rm d}s,
\end{align}
and obtain the quantum Fisher 
information matrix from Eq.~\eqref{eq:Qpure1}.

\bibliography{sample}

\section*{Acknowledgements}
This work is supported by JSPS KAKENHI Grant Number
23K13025.

\section*{Author contributions statement}

The sole author carried out all the calculations
and wrote the manuscript.

\section*{Additional information}

\textbf{Accession codes} The computational code is available at:
https://github.com/echkon/Stochastic-Parameter-shift-rule \\
\textbf{Competing interests:} 
The author declares no competing interests.

\end{document}